\documentstyle[12pt]{article}

\begin{document}
\begin{titlepage}
\title { Masses of multiquark droplets}
\author{{L. Satpathy, P.K. Sahu and V.S. Uma Maheswari} \\
Institute of Physics, Bhubaneswar-751005, India}
\maketitle
\begin{abstract}
{The mass formulae for finite lumps of strange quark matter with $u$,
$d$ and $s$ quarks, and non-strange quark matter consisting of $u$
and $d$ quarks are derived in a non-relativistic potential model. The
finite-size effects comprising the surface, curvature and even, the
Gauss curvature were consistently obtained, which shows a converging
trend. It is found that there is a possibility for the formation of
metastable strangelets of large mass. The model predicts low charge
to mass ratio as the characteristic signature of strange matter in
agreement with the relativistic studies. This study also yields an
independent estimate for the bag energy density $B$, which is in agreement
with the M.I.T bag model value.}
\end{abstract}
\smallskip
\end{titlepage}

\newpage

\noindent {\bf I. INTRODUCTION}
\vskip 1.5 true cm
The possibility of strange quark matter to be the ground state of QCD
and the probable
stability of such matter in finite lumps  called as strangelets,
speculated by many authors[1-5]
has generated a
spurt of activity during the
last few years. The general framework adopted for such studies has
been the relativistic Fermi gas model with relevant QCD
parameters like the bag constant $B$, strong coupling constant
$\alpha_c$ and the strange quark mass $m_s$. Due to large uncertainty in
these parameters, it has not yet been possible to arrive at a conclusive
result.

The calculations of Chin and Kerman \cite{chin} and Freedman and
Mclerran \cite{freedman}, which
includes the lowest order quark-quark interaction shows that infinite
quark matter even with strangeness is unstable against the nucleon
decay. Regarding the probable stability of a finite lump of quark
matter, Farhi and Jaffe \cite{farhi} were the first to consider the mass of the
strangelets by adding the surface energy and the Coulomb energy
terms to the volume energy, by initially treating the surface
tension coefficient as a parameter. Later Berger
and Jaffe \cite{berger}
developed a formula to calculate the surface tension coefficient. They
showed that surface energy is zero for massless case, and to be about
100 $MeV$ for the massive case $[m_u=m_d=0, m_s\simeq 150 MeV]$. In this
work, Berger and Jaffe have dropped the quark-quark
interaction with the supposition that it can be absorbed into the bag
constant parameter. Thus, they have developed a mass formula assuming
 a system of non-interacting quarks confined in a finite-size cavity for which
they included the surface and the Coulomb effects.

Recently, following Mardor and Svetitsky  \cite{mardor} who
have pointed out that
the curvature energy might play a decisive role for the stability of
strangelets, Madsen  \cite{madsen} has obtained an estimate for the
curvature energy
coefficient to be about 435 $MeV$ for the
massless case. This has dramatic
consequences for the stability of low mass strangelets. In view of
its importance, the curvature coefficient should also be determined
for the massive case, which has not yet been derived.
The indication of predominance of the curvature over the surface term
is already intriguing. Therefore, it is highly desirable to
develop a mass formula with surface, curvature and Gauss curvature terms
correctly calculated showing convergence trend. During the last two
decades, extensive application of non-relativistic potential
models  \cite{NR} to
the description of ground-state, excited-state and spectroscopic
properties of baryons and mesons have been amazingly successful. In
view of this, and due to the momentous nature of the problem of
stability of strangelets and its far reaching consequences, it is
worthwhile to attempt at a development of a mass formula based on a
non-relativistic approach.
In this paper, we report on such an attempt, which has the advantage of
reliable estimation of the various coefficients with practically no
parameter except the baryon number density.

In Sec.II, we present the details of our non-relativistic
formalism used to derive the mass formulae for multiquark droplets
including the finite-size effects. We then discuss the results obtained in our
model in Sec.III. Finally, we summarize and
present our conclusions in Sec.IV.

\noindent  {\bf II. FORMALISM}
\vskip 1.5 true cm

In this work, similar to Berger and Jaffe \cite{berger}
relativistic study, we
consider a system of finite number of non-interacting particles
confined in a box. This can be represented by an infinite square well
potential for which the correct enumeration of quantum states has been
done by Hill and Wheeler \cite{hill} who obtained the density of
states for such a system as
\begin{equation}
dN = {{gV}\over{2 \pi^2}}k^2dk - {{gS}\over{8\pi}}kdk +
{{gL}\over{8\pi}}dk
\end{equation}
where g is the degeneracy factor. For a spherical box of radius $R$;
$V={4\over 3}\pi R^3$, $S=4\pi R^2$ and $L=2\pi R$. It is then straight
forward to obtain an expression for the total number of particles $N$
as
\begin{equation}
N = \int_{0}^{k_F}{dN\over dk}dk = {{g V}\over{6 \pi^2}}k_F^3 -
{{g S}\over{16\pi}}k_F^{2} + {{g L}\over{8\pi}}k_F
\end{equation}
and for the total kinetic energy $E_{kin}$ as
\begin{equation}
E_{kin} = \int_{0}^{k_F}{{\hbar^2 k^2}\over{2m}}{dN\over dk}dk  =
 {{\hbar^2}\over{2m}} \left [
{{g V}\over{10 \pi^2}}k_F^5 - {{g S}\over{32\pi}}k_F^{4} +
{{g L}\over{24\pi}}k_F^3 \right ]
\end{equation}
where $k_F$ is the Fermi momentum and $m$ is the mass of the particles.
Considering a system of one kind of quarks, we have $g = 6$. We then
express the Fermi momentum $k_F$ in terms of the radius $R$, by
inverting Eq.(2) and retaining terms upto $O(N^{-1})$ as
\begin{equation}
k_{F} \simeq (\pi^2\rho)^{1/3}[1 + C_1 (\pi^2\rho)^{-1/3} + C_2
(\pi^2\rho)^{-2/3} + C_3 (\pi^2\rho)^{-1}]^{1/3}
\end{equation}
where $C_1=3\pi S/(8V)$, $C_2=2C_1^2/3 - 3\pi L /(4V)$,
$C_3=C_1^3/3 - C_1 3\pi L /(4V)$, and the total particle number
density $\rho=N/V$. Then, substituting the above expression for $k_F$
in Eq.(3), and grouping all the terms with same powers of $N$, we
arrive at the following expression for $E_{kin}$ upto the
order of $N^{0}$
\begin{equation}
E_{kin} = {{\hbar^2}\over{2m}}[{{3\pi^{4/3}}\over{5}}N
+({3\over 4})^{5/3}\pi^2 N^{2/3} + ({9\pi^3\over 16} - \pi^2)
({3\over 4\pi})^{1/3}
N^{1/3} + {9\pi^{4/3}\over{16}}({5\pi^2\over 8} - 2\pi)]\rho^{2/3}
\end{equation}
The coefficient of $N$, $N^{2/3}$, $N^{1/3}$ and $N^{0}$ are termed as volume,
surface, curvature and Gauss curvature contributions respectively to
the total kinetic energy.

Now, in order to obtain an expression for the total mass of a
multiquark system, the following important effects are to be
included.

i) For the system to be stable at finite density, pressure has to be
balanced at the surface. To ensure this, we have to add an extra term
$B V$ which is quite analogous to the MIT bag term  \cite{chodos}.

ii) The center of mass correction for a many-particle system can be
taken into account in an approximate way by multiplying the factor
$\hbar^2/2m$ in the kinetic energy expression by a factor
(${{N-1}\over N}$), as customarily done in the non-relativistic
Hartee-Fock calculations  \cite{vautherin}.

iii) In addition to the finite-size effects, the Coulomb effect has
to be included . For a spherical system of radius $R$ and charge $Ze$,
the Coulomb energy is given by $E_{Coul} = {3Z^2e^2\over{5 R}}$.

iv) Lastly, the rest mass energy of all the particles $(=mN)$ confined
in the spherical box must be included.

Thus, including all the above effects, the total mass of a $N$ particle
system is given in a compact form as
\begin{equation}
M(N) = E_{kin} + B V + E_{Coul} + m N .
\end{equation}
We then eliminate the bag energy density $B$ by using the stability
condition, which is nothing but the pressure balance equation
\begin{equation}
{d(M/N)\over{d\rho}}\mid_{\rho_{oq}} = 0.
\end{equation}
where $\rho_{oq}$ is the equilibrium quark number density.
 This gives,
 \begin{equation}
B = {2 \over 3}{E_{kin}\over{ N}} + {1\over 3}{E_{Coul}\over{ N}}
\end{equation}
 Then, in the bulk limit, neglecting the Coulomb effects, we can
rewrite $B$ as
\begin{equation}
B = {2\over 3}[{\hbar^2 \over 2m}{3 \pi^{4/3}\over{ 5}}]\rho_{oq}^{2/3}.
\end{equation}
Thus, we arrive at a formula for the bag energy density $B$ which can
be computed using
the mass of the constituent quarks and the number density of the
system.

It is interesting to recall here that the usual bag models of
hadrons employs a bag constant $B$, which is considered an
universal constant of nature. It is being widely used in the study of
hadron properties and the phase transition of hadronic matter to
quark-gluon-plasma. Normally, the value of $B$ has been determined
from the ground state properties of hadrons, which lies between 58 $MeV
fm^{-3}$ $[B^{1/4}=145 MeV]$ and $ 86 MeV fm^{-3}$ $[B^{1/4}=160 MeV]$.

On the nonrelativistic front, potential model studies have also been
very successful in describing the baryon properties.
Therefore, the above formula for $B$ given by Eq.(9) is most likely
to give an independent realistic estimate of the same.
We then express the quark number density $\rho_{oq}$ in terms
of the baryon number density $\rho_o$ as $\rho_{o} = \rho_{oq}/3$ and
fix the mass of the $u$-quark to be 300 $MeV$ which is normally used in
non-relativistic models  \cite{bhaduri}.We take the value of $\rho_o$
to be about 1.3$\rho_{nuc}$,  where the
nuclear matter density $\rho_{nuc}$ is given by 0.153 $fm^{-3}$  \cite{moller}.
Using these values, we obtain our value
of $B$ to be
about 83 $MeV fm^{-3}$. Amazingly, this value lies within the range of
relativistic values of about 58 $MeV fm^{-3} $ and  86 $MeV fm^{-3}$.
This result gives us confidence in the reliability of our model for
the study of stability of multiquark droplets.
Further, the above formula given by Eq.(9), clearly brings out the
dependence of the bag energy density $B$ on the baryon number density
and the quark mass.

Finally, we now obtain the following mass formula for a system of N
particles, each of mass $m$ and charge $q$, by
substituting the value of $B$ given by Eq.(8) in Eq.(6).
\begin{equation}
M(N) = {5\over 3} E_{kin} +{4\over 3} E_{Coul} + m N
\end{equation}

\centerline{\bf A. Two flavour system. }

The mass formula given by Eq.(10) is for a system consisting of one
kind of quarks. It can be generalised to obtain the mass of a system
consisting of quarks of more than one flavour. Here, we apply it to a
multiquark droplet consisting of $u$ and $d$ quarks, hereafter
referred to as ``udilet''.
For a given number of $u$ and $d$ quarks $N_u$
and $N_d$ respectively, we define an asymmetry parameter $\delta =
(N_d - N_u)/N_q$, where $N_q = N_u + N_d$. So, the quark density for
each flavour can be expressed in terms of $\delta$ and $\rho_{oq}$ as
\begin{eqnarray}
\rho_u = {\rho_{oq}\over 2}(1-\delta) \nonumber \\
\rho_d = {\rho_{oq}\over 2}(1+\delta)
\end{eqnarray}
with $\rho_{oq} = N_q/V$.
The total kinetic energy for a two flavour system is obtained by
summing over the respective contributions, given by Eq.(5) for $u$
and $d$ quarks. Then, one can obtain the total mass of an
udilet in terms of the baryon number density $\rho_o [=\rho_{oq}/3]$
using Eqs.(10) and (11) as
\begin{eqnarray}
M_2(A)& = &{5\over 3}{\hbar^2\over{2m_u}}{3 \pi^{4/3}\over 5}
[\rho_u^{5/3} + \rho_d^{5/3}]{4\pi R^3\over 3} \nonumber \\
&+ &{5\over 3}{\hbar^2\over{2m_u}}{3 \pi^{5/3}\over 16}
[\rho_u^{4/3} + \rho_d^{4/3}]{4\pi R^2}  \nonumber \\
&+& {5\over 3}{\hbar^2\over{2m_u}}[{9 \pi^2\over 32} - {\pi\over 2}]
[\rho_u + \rho_d]{2\pi R} \nonumber  \\
&+& {5\over 3}{\hbar^2\over{2m_u}}{9 \pi^{4/3}\over 16}[{5 \pi^2\over 8}
- 2\pi] [\rho_u^{2/3} + \rho_d^{2/3}]\nonumber  \\
&+& {4\over 3}[{3 Z e^2\over{5 R}}] + 3 m_u A
\end{eqnarray}
where we have used $m_d=m_u$ and $A=N_q/3$.
By substituting for $\rho_u$, $\rho_d$ and the radius $
R = r_o A^{1/3}$ with  ${4\over 3}\pi r_o^3 \rho_o = 1$, we thus arrive
at the Bethe-Weizsacker like mass formula for the udilets upto the
order of $A^{-1}$ and $\delta^{2}$.
\begin{equation}
{M_2(A)\over A} = a_{v2} + a_{s2} A^{-1/3} + a_{c2} A^{-2/3} + a_{o2}
A^{-1} + a_{z} Z^2 A^{-4/3} + 3 m_u
\end{equation}
where $a_{v2}, a_{s2}$ etc are defined as $a_{v2} = a_v(1 + {5\over 9}
\delta^2)$, $a_{s2} = a_s(1 + {2\over 9}\delta^2)$, $a_{c2} = a_c$,
$a_{o2} = a_o(1 - {1\over 9}\delta^2)$, and the various coefficients
are given by,
\begin{eqnarray}
a_v & = &{5\over 3}{\hbar^2\over{2m_u}}{9 \pi^{4/3}\over
5}(1.5)^{2/3} \rho_o^{2/3}\nonumber \\
a_s & = &{5\over 3}{\hbar^2\over{2m_u}}({3 \over 4})^{5/3}({3 \over
2})^{4/3}2 \pi^2 \rho_o^{2/3} \nonumber \\
a_c &=& {5\over 3}{\hbar^2\over{2m_u}}[{9 \pi\over 16} - 1]
3 \pi^2 ({3\over{4\pi}})^{1/3}(1.5)^{2/3} \rho_o^{2/3}  \nonumber  \\
a_o&=& {5\over 3}{\hbar^2\over{2m_u}}[{9 \pi^{4/3}\over 8}[{5 \pi^2\over 8}
- 2\pi]- {6 \pi^{4/3}\over 5}] \rho_o^{2/3} \nonumber  \\
a_z&=& {4\over 3}[{3 e^2\over{5 r_0}}]
\end{eqnarray}
\vfill
\newpage
\centerline{\bf B. Three flavour system }

Similar to the two flavour system discussed above , we now consider a
multiquark droplet
consisting of $u$, $d$ and $s$ quarks usually referred to as
strangelet. For this case of three flavour system, we need to define
two asymmetry parameters namely, $\delta_{ud} = (N_d - N_u)/N_q$ and
$\delta_{us} = (N_s - N_u)/N_q$, where $N_q = N_u + N_d + N_s$. We
can then express the quark density of each flavour in terms of
$\delta_{ud}, \delta_{us}$ and $\rho_{oq}$ as
\begin{eqnarray}
\rho_u &=& {\rho_{oq}\over 3}(1 - \delta_{us} - \delta_{ud})\nonumber\\
\rho_d &=& {\rho_{oq}\over 3}(1 - \delta_{us} + 2\delta_{ud})\nonumber\\
\rho_s &=& {\rho_{oq}\over 3}(1 +2\delta_{us} - \delta_{ud}).
\end{eqnarray}
The total kinetic energy for this system is obtained by summing over
the respective contributions given by Eq.(5) for $u$, $d$ and $s$ quarks.
Using Eqs.(10) and (15), we thus arrive at the following mass formula
for strangelets in terms of the baryon number $A$ and its density
$\rho_{o}$ retaining, terms upto $O(A^{-1})$ and second order in the
asymmetry parameters.
\begin{equation}
{M_3(A)\over A} = a_{v3} + a_{s3} A^{-1/3} + a_{c3} A^{-2/3} + a_{o3}
A^{-1} + a_{z} Z^2 A^{-4/3} + (f_u + f_d) m_u + f_s m_s
\end{equation}
where the quark fractions can be expressed using Eq.(15), as follows .
 \begin{eqnarray}
f_u &=&{N_u\over A} = 1 - \delta_{us} - \delta_{ud}\nonumber\\
f_d &=& {N_d\over A} = 1 - \delta_{us} + 2\delta_{ud}\nonumber\\
f_s &=& {N_s\over A} = 1 +2\delta_{us} - \delta_{ud}
\end{eqnarray}
The various coefficients $a_{v3}, a_{s3}$ ${\it etc}$ in Eq.(16) are
given by,
\begin{eqnarray}
a_{v3} & = &{5\over 3}c_v [ ({2\over m_u} +{1\over m_s})+({1\over m_u} -
{1\over m_s}){5\over 3} (\delta_{ud}-2\delta_{us})+{5\over
9}({\Delta_1^{2}\over m_u}+{\Delta_2^{2}\over m_s})] \nonumber \\
a_{s3} & = &{5\over 3}c_s [ ({2\over m_u} +{1\over m_s})+({1\over m_u} -
{1\over m_s}){4\over 3} (\delta_{ud}-2\delta_{us})+{2\over
9}({\Delta_1^{2}\over m_u}+{\Delta_2^{2}\over m_s})] \nonumber \\
a_{c3} & = &{5\over 3}c_c [ ({2\over m_u} +{1\over m_s})+({1\over m_u} -
{1\over m_s})(\delta_{ud}-2\delta_{us})]\nonumber \\
a_{o3} & = &{5\over 3}c_o [ ({2\over m_u} +{1\over m_s})+({1\over m_u} -
{1\over m_s}){2\over 3} (\delta_{ud}-2\delta_{us})-{1\over
9}({\Delta_1^{2}\over m_u}+{\Delta_2^{2}\over m_s})] \nonumber \\
\end{eqnarray}
where
\begin{eqnarray}
c_v & = &{\hbar^2\over 2}{3\over 5}\pi^{4/3}\rho_o^{2/3}\nonumber \\
c_s & = &{\hbar^2\over 2}({3 \over 4})^{5/3}\pi^2 \rho_o^{2/3} \nonumber \\
c_c &=& {\hbar^2\over 2}[{9 \pi^3\over 16} -
\pi^2]({3\over{4\pi}})^{1/3} \rho_o^{2/3}  \nonumber  \\
c_o&=& {\hbar^2\over 2}[{9 \pi^{4/3}\over 16}({5 \pi^2\over 8}
- 2\pi)- {3 \pi^{4/3}\over 5}] \rho_o^{2/3} \nonumber \\
\Delta_1^{2}&=&{(\delta_{ud}+\delta_{us})}^{2}
+ {(2 \delta_{ud}-\delta_{us})}^{2}\nonumber \\
\Delta_2^{2}&=&{(2 \delta_{us}-\delta_{ud})}^{2}.
\end{eqnarray}

\noindent {\bf III. RESULTS AND DISCUSSIONS.}

The mass formulae given by Eqs.(13) and (16) are most general, which
can be used for any arbitrary combination of $u$,$d$ and $s$ quarks.
To calculate the various coefficients like volume, surface ${\it etc}$ ,
we have chosen the mass of $u$ and $d$ quarks to be 300 $MeV$ which
lies within the range of values
normally used in the non-relativistic potential model calculations
 \cite{bhaduri} of baryon spectroscopy. We take the mass
of $s$ quark to be 400 $MeV$, which is quite reasonable.
The baryon number density $\rho_o$ is assumed to be about 1.3 times the nuclear
matter density $\rho_{nuc}=0.153 fm^{-3}$.
For the sake of convenience, we
calculate these coefficients by imposing charge neutrality
condition, which for udilets implies $N_{d}= 2N_{u}$, $\delta
={1\over 3}$ and for strangelets $N_{u}=N_{d}=N_{s}$,
$\delta_{ ud}=0$,$\delta_{ us}=0$.
The value of the coefficients so computed
for udilets and strangelets are given in Table I. It is satisfying to
note that the contribution of the various terms successively
decreases leading to a faster convergence of $M(A)/A$. It is
interesting to note that the Gauss curvature term has negative sign
and so has the opposite effect on the mass compared to surface and
curvature terms. This is in fact is a welcome feature. It must be
realised that these kind of mass formulae based on Fermi gas
approximation are not valid for very small baryon numbers. Because
of its general power series structure in $A^{-1/3}$, the higher order
contributions becomes progressively important for low $A$ values.
However, we feel these formulae are more appropriate for baryon
number $A > 50$ as also noted by Jaffe  \cite{jaffe}. In Table II,
we have
presented the corresponding masses of udilets and strangelets for
certain typical values of $A$. From symmetry considerations,
strangelets are expected to be more stable than the corresponding
udilets for a given baryon number.This is borne out quite well in our
calculations as can be seen from Table II.
The energy per baryon for strange quark matter is about 1275
$MeV$, and for non-strange quark matter consisting of $u$ and $d$
quarks , it is about 1320 $MeV$. Thus, both
these quark matters are unstable against nucleon decay. So, we find in
this model the nuclear matter to be the ground state of hadronic
matter. However, this does not ${\it a priori}$ preclude the possibility of
meta-stable strangelet formation.
Before, we embark upon the investigation of the possibility of
strangelet formation in this model, we would like to consider the
influence of the baryon number density $\rho_o$ and strangeness on
the relative stability of strangelets and udilets.

In Fig.1, we illustrate the effect of $\rho_o$ on the relative
stability between the strangelets and udilets, using the general mass
formula for strangelets and udilets given by Eqs.(16) and (13). We
 have found the most flavour-stable isobar, by minimising the mass
with respect
to $\delta$ for udilets, and $\delta_{ud}$ and $\delta_{us}$ for
strangelets. In Fig.1a, the difference between the masses of the
strangelets and udilets so obtained are plotted as a function of
$\rho_o / \rho_{nuc}$, for a typical baryon number $A=400$. It can be
seen that mass of the strangelets is lower than the
corresponding udilets for all densities. The minimum difference
between their masses is about 4 MeV. This feature is true for any
baryon number $A$ is shown in Fig.1b, where the surface and Coulomb
effects are neglected

In Fig.2, we illustrate the effect of strangeness on the relative
stability between the strangelets and utilets for the same $A=400$.
Using the general formula for strangelets given by Eq.(16), we
have determined for a given $\delta_{us}$, the flavour-stable value of mass
$M_3(A)$ by minimizing the toal mass with respect to $\delta_{ud}$.
For two chosen
values of $\rho_o$, namely 1.27 $\rho_{nuc}$ and 2.14 $\rho_{nuc}$,
the results so obtained are presented as solid curves in Fig.2 as a
function of $f_{s}$, where $f_{s}$ in terms of $\delta_{ud}$ and
$\delta_{us}$ is given by Eq.(17). It is
interesting to find that the mass shows a well defined minimum for $f_{s}$
at about 0.7.
Then to show the relative stability with respect to udilets having
the same baryon number $A$, we have minimised the mass $M_2(A)$ given
by Eq.(13) with respect to $\delta$.
We have represented the most stable mass of udilets so obtained for the
same two values of $\rho_o$ as dashed lines. It can be seen from Fig.2
that at the optimum value of $f_{s}$, the difference between the mass of
strangelets and udilets is maximum, which gradually decreases and
tends to zero as $f_{s}$ is varied about the optimum value.
Therefore, there is a maximum value of $f_{s}$, referred to as
$f_{s}^{m}$, above which udilets becomes more stable than the strangelets.
Thus, there is a `window' in $f_{s}$
space, only within which strangelets are more stable than the
corresponding udilets. This window region depends upon density and
increases as $\rho_o$ is increased, which implies that hyper
strange multiquark droplets are more probable at higher densities.
This is shown in Fig.3, where $f_{s}^{m}$ is plotted as a function of
$\rho_{o} / \rho_{nuc}$ for the same value of $A$, namely $A =
400$.
It can be seen that as the density increases, $f_{s}^{m}$ increases.
The shaded portion in Fig.3 represents the region where strangelets are more
stable relative to udilets, and the unshaded region represents the vice-versa.

Now, we would like to investigate the probable stability of
strangelets in this model.
For a given $\delta_{us}$, we determine the most stable mass per
baryon $M_V(A)/A$ of the strangelet in the bulk limit, by minimizing
the mass given by Eq.(16) with respect to $\delta_{ud}$.
In the absence of Coulomb interaction $N_{u} = N_{d}$; $f_{s}$($=N_{s}/A$)
is determined by using these values of
$\delta_{ud}$ and $\delta_{us}$ in Eq.(17).
Following Chin and Kerman, we plot in Fig.4  ${M_V/A}-M_N$
versus $f_{s}$, where the masses of the nucleon, lambda,
cascade and omega are shown as solid dots. It can be seen that
strangelets are unstable against nucleon and lambda decay even at
the optimum value of $f_{s}$, which is about 0.7.
As the optimum value of $f_{s}$ is
less than unity, the number of $s$ quarks is less
than the number of $u$ and $d$ quarks.
Therefore, the excess of the $u$ and $d$ quarks over that of $s$
quarks, will occupy the higher momentum states than the $s$
quarks. These
$u$ and $d$ quarks will then form nucleons and strongly decay, analogous to the
nucleon dripping from a highly radioactive nucleus.
At this stage, as the mass of the lambda is higher than that of  the
nucleon, we expect the nucleon decay
to be more probable than the lambda particle. Thus, the system is
likely to be driven towards a state of greater strangeness, where
it is stable against cascade and omega decay, as per
our calculations.
Under these conditions, strangelets can be formed in a metastable
state. It can only decay by weak leptonic decay process. This possibility
gives us some hope of detecting strangelets, but only of very large mass.

In the present model, we have the density as the only  free
parameter, because the
masses of $u$, $d$ and $s$ quarks are constrained by the non-relativistic
potential model studies of baryon spectroscopy. We find that with the
increase of density, the masses of the quarks have to be lowered in
order to make the system stable against omega emission. For a
given density, the masses of these quarks can be varied within the
acceptable limits to ensure the same.
In Fig.5, fixing the density at about 1.3 $\rho_{nuc}$, we have
shown the range of $m_u$ and $m_s-m_u$ for which the omega baryon is
bound in our model. It is interesting to note that the mass of $s$
quark comes about
to be order of 100 $MeV$ higher than the $u$ and $d$ quark masses,
which is quite reasonable.
Thus, we find that there is a sizable range of these parameters in which
metastable strangelet formation is allowed.

As found by Farhi and Jaffe  \cite {farhi} in the relativistic study
of strange matter, a
possible signature for strange matter is its characteristic low
charge to mass ratio, unlike in the case of ordinary nuclei, where
$Z\simeq A/2$. We would like to investigate whether our model would
corroborate their finding. So, for the case of a strangelet of baryon
number $A$, we obtain the
flavour-stable value of the charge number $Z^{*}_{3}$ given by
$$Z^*_3=-A[\delta^*_{us}+\delta^*_{ud}]$$
where $\delta^*_{us}$ and $\delta^*_{ud}$ are determined by
simultaneously the two equations,
$${dM_3/A\over d\delta_{us}}\mid_{\delta^*_{us}}=0$$
$${dM_3/A\over d\delta_{ud}}\mid_{\delta^*_{ud}}=0$$
using Eq.(16).

Similarly, the flavour-stable value of the charge number $Z^*_2$ for
the corresponding udilet is obtained using
$$Z^*_2= {A\over 2}[1-3\delta^*].$$
with $\delta^*$ determined from the condition
$${dM_2/A\over d\delta}\mid_{\delta^*}=0$$
where, the mass of the utilets $M_2(A)$ is given by Eq.(13).

In Table III, we have presented the values of $Z^*_2$ and $Z^*_3$
obtained in the above calculation for various baryon number $A$.
For the sake of comparison we have also given the value $Z = A/2$,
which is appropriate for nuclei of large $A$. It can be seen that the
$Z_{min}$ for the udilets $(Z^*_2)$ is somewhat lower than the
corresponding nuclei, which becomes substantial with the increase of
$A$. However, for the strangelets the $Z_{min}$ value $(Z^*_3)$ is
strikingly low. It is indeed satisfying to find that the present
non-relativistic study arrives at a similar conclusion as that of the
relativistic study of Farhi and Jaffe. Thus, the low charge to mass
ratio as a strong signature of strangelets is being reinforced by
our present study.

In summarising our results and discussions, we find that the present
non-relativistic model has surprisingly succeeded in describing most
of the features of strangelets as found in the relativistic studies.
This success and the reliability of our model can be solely
attributed to the sound foundation based on the following two
elements, due to a which straight forward calculation was possible:

(i). The confinement of quarks is best represented by the infinite
square well potential.

(ii). For this potential, the exact enumeration of quantum
states was given by Hill and Wheeler, which has been used.
\vfill

\newpage
\noindent {\bf IV. CONCLUSION }

In conclusion, we have developed the mass formulae for finite lump of
quark matter for systems with $u$ and $d$ quarks, and also for
systems with $u$ , $d$ and $s$ quarks, in a non-relativistic
model. The complete and consistent mass formula including all
finite-size effects comprising the surface, curvature, and even the
Gauss curvature has been derived for both the strangelets and udilets.
For the first time, we have obtained an estimate of the Gauss
curvature term.
Further, this model study provides an independent estimate of the bag energy
density $B$, which we find to be in agreement with the M.I.T bag
model value.

We have shown that for a given baryon density there exists an upper
bound on the strangeness below which strangelets are more
stable than the udilets. This upper limit on the number of $s$ quarks
increases as density increases, which implies that hyperstrange
strangelets are more probable at higher densities.

Finally, our calculation shows that there is a finite possibility
for the formation of metastable strangelets of very large mass.
The present study also predicts low charge to mass ratio as the
characteristic signature of strange matter in agreement with the
relativistic studies.

The success of this model is due to the use of infinite square well
potential, which adequately describes the confinement of quarks and
for which exact quantum density of states can be obtained.

\vfill
\newpage
\centerline {\bf FIGURE CAPTIONS}
\noindent {\bf FIG.1}. The difference between the mass per
baryon of strangelets
and udilets is plotted against the baryon number density
$\rho_o$ in the units of the nuclear matter density
$\rho_{nuc}$; (a) for $A=400$ and
(b) in the bulk limit neglecting the Coulomb effects.

\noindent {\bf FIG.2}.The mass per baryon of strangelets
relative to the nucleon mass is plotted
against the $s$ quark fraction $f_{s}$ as solid curves for two
chosen values of baryon density $\rho_{o}$, for a given baryon
number $A=400$. The upper curve corresponds to $\rho_{o}=2.14
\rho_{nuc}$ and the lower one to 1.27$\rho_{nuc}$. The
corresponding mass of udilets
relative to the nucleon mass is shown as dashed lines.

\noindent {\bf FIG.3}.The maximum value of the $s$ quark
fraction $f_{s}^{m}$ is
plotted against the baryon density $\rho_{o}$ in units of the  nuclear
matter density $\rho_{nuc}$, for a given baryon number $A=400$. The
dotted region represents  where the strangelets are more stable than
the corresponding udilets.

\noindent {\bf FIG.4}. The mass per baryon of the strangelets
$M_{v}/A$ relative
to the nucleon
mass in the bulk limit (neglecting the Coulomb effects) is plotted
against the $s$ quark fraction $f_{s}$. The masses of the nucleon
$(M_{N})$, lambda particle $(M_{\Lambda})$, cascade particle $(M_{\Xi})$
and omega particle $(M_{\Omega})$ are shown as solid dots.

\noindent {\bf FIG.5}. The range of the parameters $m_u$ and
$m_s$, the mass of $u$ and $s$ quarks respectivley favouring
stability of strange quark mass against omega emission. The
contour plot of the mass per baryon
$M_{v}/A$ of the strangelets with $f_{s}=3$ (see Eq.17)
in the bulk limit (neglecting the Coulomb effects) is plotted against
$m_{u}$ and difference between $m_{u}$ and $m_{s}$. The
corresponding numerical values of
$M_{v}/A$ are shown .
\vfill

\newpage
\centerline {\bf TABLE CAPTIONS}

\noindent {\bf Table I}. The values of the volume, surface,
curvature and Gauss curvature coefficients,
$a_{v}$, $a_{s}$, $a_{c}$ and $a_{o}$ respectively
obtained using
the charge neutrality condition are given for both the udilets and
strangelets.

\noindent {\bf Table II}. The mass per baryon for strangelets $(M_3/A)$
and udilets $(M_2/A)$ calculated using the charge neutrality condition are
given for various baryon number $A$.

\noindent {\bf Table III}. The flavour-stable value of the charge number for
udilets $(Z_2^*)$ and strangelets $(Z_3^*)$ are given for various baryon
number $A$.
\vfill
\newpage
\centerline {\bf Table I}
\vspace {0.5in}
\begin{center}
\begin{tabular}{|c|c|c|c|c|}
\hline
\multicolumn{1}{|c|}{$System$} &
\multicolumn{1}{|c|}{$a_v$} &
\multicolumn{1}{|c|}{$a_s$} &
\multicolumn{1}{|c|}{$a_c$} &
\multicolumn{1}{|c|}{$a_o$} \\
\multicolumn{1}{|c|}{} &
\multicolumn{1}{|c|}{ (MeV)} &
\multicolumn{1}{|c|}{ (MeV)} &
\multicolumn{1}{|c|}{(MeV)} &
\multicolumn{1}{|c|}{(MeV)} \\
\hline
udilet&417.53&779.14&510.59&-286.77\\
(udd) &      &      &      &\\
strangelet&275.10&608.89&468.04&-304.67\\
(uds)    &      &      &      &\\
\hline
\end{tabular}
\end{center}
\vfill

\newpage
\centerline {\bf Table II}
\vspace {0.5in}
\begin{center}
\begin{tabular}{|c|c|c|}
\hline
\multicolumn{1}{|c|}{$A$} &
\multicolumn{1}{|c|}{$M_2/A$} &
\multicolumn{1}{|c|}{$M_3/A$} \\
\multicolumn{1}{|c|}{} &
\multicolumn{1}{|c|}{$(MeV)$} &
\multicolumn{1}{|c|}{$(MeV)$} \\
\hline
50.00&1560.90&1468.77\\
100.00&1506.22&1424.96\\
200.00&1464.25&1391.38\\
300.00&1444.35&1375.48\\
500.00&1423.22&1358.63\\
1000.00&1400.26&1340.36\\
3000.00&1373.91&1319.46\\
5000.00&1364.78&1312.25\\
10000.00&1354.76&1304.34\\
\hline
\end{tabular}
\end{center}
\vfill

\newpage
\centerline {\bf Table III}
\vspace {0.5in}
\begin{center}
\begin{tabular}{|c|c|c|c|}
\hline
\multicolumn{1}{|c|}{$A$} &
\multicolumn{1}{|c|}{$Z=A/2$} &
\multicolumn{1}{|c|}{$Z_2^*$} &
\multicolumn{1}{|c|}{$Z_3^*$} \\
\hline
300.00&150.00&103.79&28.78\\
500.00&250.00&152.78&44.27\\
1000.00&500.00&246.42&74.38\\
3000.00&1500.00&470.38&147.40\\
5000.00&2500.00&609.49&193.17\\
10000.00&5000.00&837.78&268.86\\
50000.00&25000.00&1588.38&521.26\\
100000.00&50000.00&2041.08&674.98\\
1000000.00&500000.00&4503.32&1516.66\\
\hline
\end{tabular}
\end{center}

\end{document}